%%%%%%%%%%%%%%%%%%%%%%%%%%%%%%%%%%%%%%%%%%%%%%%%%%%%%%%%%%%%%%%%%%
%%                                                              %%
%%                                                              %%
%%                Plain TeX file for the paper:                 %%
%%                                                              %%
%% "Supersymmetric Quantum Mechanics with a Point Interaction"  %%
%%                                                              %%
%%                             by                               %%
%%                   T. Uchino and I. Tsutsui                   %%
%%                                                              %%
%%                                                              %%
%%%%%%%%%%%%%%%%%%%%%%%%%% October 2002 %%%%%%%%%%%%%%%%%%%%%%%%%%

%%%%%%%%%%%%%%%%%%%%%%%%%%%%%%%%%%%%%%%%%%%%%%%%%%%%%%
%%%%%%%%    Collection of macro's for use    %%%%%%%%%
%%%%%%%%    with plain tex                   %%%%%%%%%
%%%%%%%%%%%%%%%%%%%%%%%%%%%%%%%%%%%%%%%%%%%%%%%%%%%%%%

%%% First some fonts %%%%%%%%%%%%%%%%%%%%%%%%%%%%%%%%%
\font\bigbold=cmbx12
\font\ninerm=cmr9      \font\eightrm=cmr8    \font\sixrm=cmr6
\font\fiverm=cmr5
\font\ninebf=cmbx9     \font\eightbf=cmbx8   \font\sixbf=cmbx6
\font\fivebf=cmbx5
\font\ninei=cmmi9      \skewchar\ninei='177  \font\eighti=cmmi8
\skewchar\eighti='177  \font\sixi=cmmi6      \skewchar\sixi='177
\font\fivei=cmmi5
\font\ninesy=cmsy9     \skewchar\ninesy='60  \font\eightsy=cmsy8
\skewchar\eightsy='60  \font\sixsy=cmsy6     \skewchar\sixsy='60
\font\fivesy=cmsy5     \font\nineit=cmti9    \font\eightit=cmti8
\font\ninesl=cmsl9     \font\eightsl=cmsl8
\font\ninett=cmtt9     \font\eighttt=cmtt8
\font\tenfrak=eufm10   \font\ninefrak=eufm9  \font\eightfrak=eufm8
\font\sevenfrak=eufm7  \font\fivefrak=eufm5
\font\tenbb=msbm10     \font\ninebb=msbm9    \font\eightbb=msbm8
\font\sevenbb=msbm7    \font\fivebb=msbm5
\font\tenssf=cmss10    \font\ninessf=cmss9   \font\eightssf=cmss8
\font\tensmc=cmcsc10

%%% Some Families %%%%%%%%%%%%%%%%%%%%%%%%%%%%%%%%%%%%
\newfam\bbfam   \textfont\bbfam=\tenbb \scriptfont\bbfam=\sevenbb
\scriptscriptfont\bbfam=\fivebb  \def\Bbb{\fam\bbfam}
\newfam\frakfam  \textfont\frakfam=\tenfrak \scriptfont\frakfam=%
\sevenfrak \scriptscriptfont\frakfam=\fivefrak  \def\frak{\fam\frakfam}
\newfam\ssffam  \textfont\ssffam=\tenssf \scriptfont\ssffam=\ninessf
\scriptscriptfont\ssffam=\eightssf  
\def\smc{\tensmc}

%%% Definition of 8 point %%%%%%%%%%%%%%%%%%%%%%%%%%%%
\def\eightpoint{\textfont0=\eightrm \scriptfont0=\sixrm
\scriptscriptfont0=\fiverm  \def\rm{\fam0\eightrm}%
\textfont1=\eighti \scriptfont1=\sixi \scriptscriptfont1=\fivei
\def\oldstyle{\fam1\eighti}\textfont2=\eightsy
\scriptfont2=\sixsy \scriptscriptfont2=\fivesy
\textfont\itfam=\eightit         \def\it{\fam\itfam\eightit}%
\textfont\slfam=\eightsl         \def\sl{\fam\slfam\eightsl}%
\textfont\ttfam=\eighttt         \def\tt{\fam\ttfam\eighttt}%
\textfont\frakfam=\eightfrak     \def\frak{\fam\frakfam\eightfrak}%
\textfont\bbfam=\eightbb         \def\Bbb{\fam\bbfam\eightbb}%
\textfont\bffam=\eightbf         \scriptfont\bffam=\sixbf
\scriptscriptfont\bffam=\fivebf  \def\bf{\fam\bffam\eightbf}%
\abovedisplayskip=9pt plus 2pt minus 6pt   \belowdisplayskip=%
\abovedisplayskip  \abovedisplayshortskip=0pt plus 2pt
\belowdisplayshortskip=5pt plus2pt minus 3pt  \smallskipamount=%
2pt plus 1pt minus 1pt  \medskipamount=4pt plus 2pt minus 2pt
\bigskipamount=9pt plus4pt minus 4pt  \setbox\strutbox=%
\hbox{\vrule height 7pt depth 2pt width 0pt}%
\normalbaselineskip=9pt \normalbaselines \rm}

%%% Definition of 9 point %%%%%%%%%%%%%%%%%%%%%%%%%%%%
\def\ninepoint{\textfont0=\ninerm \scriptfont0=\sixrm
\scriptscriptfont0=\fiverm  \def\rm{\fam0\ninerm}\textfont1=\ninei
\scriptfont1=\sixi \scriptscriptfont1=\fivei \def\oldstyle%
{\fam1\ninei}\textfont2=\ninesy \scriptfont2=\sixsy
\scriptscriptfont2=\fivesy
\textfont\itfam=\nineit          \def\it{\fam\itfam\nineit}%
\textfont\slfam=\ninesl          \def\sl{\fam\slfam\ninesl}%
\textfont\ttfam=\ninett          \def\tt{\fam\ttfam\ninett}%
\textfont\frakfam=\ninefrak      \def\frak{\fam\frakfam\ninefrak}%
\textfont\bbfam=\ninebb          \def\Bbb{\fam\bbfam\ninebb}%
\textfont\bffam=\ninebf          \scriptfont\bffam=\sixbf
\scriptscriptfont\bffam=\fivebf  \def\bf{\fam\bffam\ninebf}%
\abovedisplayskip=10pt plus 2pt minus 6pt \belowdisplayskip=%
\abovedisplayskip  \abovedisplayshortskip=0pt plus 2pt
\belowdisplayshortskip=5pt plus2pt minus 3pt  \smallskipamount=%
2pt plus 1pt minus 1pt  \medskipamount=4pt plus 2pt minus 2pt
\bigskipamount=10pt plus4pt minus 4pt  \setbox\strutbox=%
\hbox{\vrule height 7pt depth 2pt width 0pt}%
\normalbaselineskip=10pt \normalbaselines \rm}

%%% Macro to generate the equation #'s automatically.
%%% To use start each new section (eg 3) with the commands
%%% \secno=3 \meqno=1 :this will start the equations with (3.1)
%%% Then in place of \eqno(3.1) type \eqn\descriptivename . To refer
%%% back to the equation simply type (\descritivename)
%%% For the appendixset \secno=0, \appno=1\meqno=1 etc
%%%
\global\newcount\secno \global\secno=0 \global\newcount\meqno
\global\meqno=1 \global\newcount\appno \global\appno=0
\newwrite\eqmac \def\romappno{\ifcase\appno\or A\or B\or C\or D\or
E\or F\or G\or H\or I\or J\or K\or L\or M\or N\or O\or P\or Q\or
R\or S\or T\or U\or V\or W\or X\or Y\or Z\fi}
\def\eqn#1{ \ifnum\secno>0 \eqno(\the\secno.\the\meqno)
\xdef#1{\the\secno.\the\meqno} \else\ifnum\appno>0
\eqno({\rm\romappno}.\the\meqno)\xdef#1{{\rm\romappno}.\the\meqno}
\else \eqno(\the\meqno)\xdef#1{\the\meqno} \fi \fi
\global\advance\meqno by1 }

%%% Macro to do the refs %%%%%%%%%%%%%%%%%%%%%%%%%%%%%
\global\newcount\refno \global\refno=1 \newwrite\reffile
\newwrite\refmac \newlinechar=`\^^J \def\ref#1#2%
{\the\refno\nref#1{#2}} \def\nref#1#2{\xdef#1{\the\refno}
\ifnum\refno=1\immediate\openout\reffile=refs.tmp\fi
\immediate\write\reffile{\noexpand\item{[\noexpand#1]\ }#2\noexpand%
\nobreak.} \immediate\write\refmac{\def\noexpand#1{\the\refno}}
\global\advance\refno by1} \def\semi{;\hfil\noexpand\break ^^J}
\def\nl{\hfil\noexpand\break ^^J} \def\refn#1#2{\nref#1{#2}}

\def\ann#1#2#3{{\it Ann.\ Phys.}\ {\bf {#1}} ({#2}) #3}

\def\plA#1#2#3{{\it Phys.\ Lett.}\ {\bf {#1}A} ({#2}) #3}

%%% Numbering does not start on title page %%%%%%%%%%%
\newif\iftitlepage \titlepagetrue \newtoks\titlepagefoot
\titlepagefoot={\hfil} \newtoks\otherpagesfoot \otherpagesfoot=%
{\hfil\tenrm\folio\hfil} \footline={\iftitlepage\the\titlepagefoot%
\global\titlepagefalse \else\the\otherpagesfoot\fi}

%%% Abstract %%%%%%%%%%%%%%%%%%%%%%%%%%%%%%%%%%%%%%%%%
\def\abstract#1{{\parindent=30pt\narrower\noindent\ninepoint\openup
2pt #1\par}}

%%% A nicer footnote (\note) %%%%%%%%%%%%%%%%%%%%%%%%%
\newcount\notenumber\notenumber=1 \def\note#1
{\unskip\footnote{$^{\the\notenumber}$} {\eightpoint\openup 1pt #1}
\global\advance\notenumber by 1}

%%% Date %%%%%%%%%%%%%%%%%%%%%%%%%%%%%%%%%%%%%%%%%%%%%
\def\today{\ifcase\month\or January\or February\or March\or
April\or May\or June\or July\or August\or September\or October\or
November\or December\fi \space\number\day, \number\year}

%%% More general stuff %%%%%%%%%%%%%%%%%%%%%%%%%%%%%%%
\def\pagewidth#1{\hsize= #1}  \def\pageheight#1{\vsize= #1}
\def\hcorrection#1{\advance\hoffset by #1}
\def\vcorrection#1{\advance\voffset by #1}

%%% Output layout of the text %%%%%%%%%%%%%%%%%%%%%%%%
\pageheight{23cm}
\pagewidth{15.7cm}
\hcorrection{-1mm}
\magnification= \magstep1
\parskip=5pt plus 1pt minus 1pt
\tolerance 8000
\def\bsk{\baselineskip= 14.5pt plus 1pt minus 1pt}
\bsk

%%% Definition of extra symbols R, C, Z, N %%%%%%%%%%%
\font\extra=cmss10 scaled \magstep0  \setbox1 = \hbox{{{\extra R}}}
\setbox2 = \hbox{{{\extra I}}}       \setbox3 = \hbox{{{\extra C}}}
\setbox4 = \hbox{{{\extra Z}}}       \setbox5 = \hbox{{{\extra N}}}

          % Actual special symbol: R

   % Actual special symbol: C

           % Actual special symbol: Z

            % Actual special symbol: N

%%% Some useful macros %%%%%%%%%%%%%%%%%%%%%%%%%%%%%%%
\def\frac#1#2{{#1\over#2}}

\def\pmb#1{\setbox0=\hbox{$#1$} \kern-.025em\copy0\kern-\wd0
    \kern.05em\copy0\kern-\wd0 \kern-.025em\raise.0433em\box0 }

\def\ve{\vfill\eject}

\def\tr{\hbox{tr}}
\def\({\left(}
\def\){\right)}

\def\R{{\Bbb R}}  
  
\def\Z{{\Bbb Z}}  
\def\C{{\Bbb C}}

\def\figone{line1.eps}
\def\figtwo{circleline.eps}
\def\figthree{spectline.eps}

\input epsf

\let\omitpictures=N

%%% References for the paper:

{

\refn\RS
{M. Reed, B. Simon,
\lq\lq Methods of Modern Mathematical Physics\rq\rq , 
{\sl Vol.II}, Academic Press, New York, 1980}

\refn\AGHH
{S. Albeverio, F. Gesztesy, R. H{\o}egh-Krohn and H. Holden,
\lq\lq Solvable Models in Quantum Mechanics\rq\rq,
Springer, New York, 1988}

\refn\FCT
{T. F\"{u}l\"{o}p, T. Cheon and I. Tsutsui,
{\it Classical Aspects of Quantum Walls in One Dimension}, KEK preprint
2001-134, quant-ph/0111057, to appear in {\it Phys.\ Rev.}\ {\bf {A}}}

\refn\CFT
{T. Cheon, T. F\"{u}l\"{o}p and I. Tsutsui,
\ann{294}{2001}{1-23}}

\refn\TFC
{I. Tsutsui, T. F\"{u}l\"{o}p and T. Cheon,
 {\it J. Phys. Soc. Jpn.} {\bf 69} (2000) 3473}

\refn\Witt
{E. Witten,
{\it Nucl.Phys.} {\bf B 185} (1981) 513}

\refn\Junker
{G. Junker,
\lq\lq Supersymmetric Methods in Quantum
and Statistical
Physics\rq\rq,
Springer,
Berlin, 1996}

\refn\AG
{N.I. Akhiezer and I.M. Glazman,
\lq\lq Theory of Linear Operators in Hilbert Space\rq\rq,
{\sl Vol.II},
Pitman Advanced Publishing Program, Boston, 1981}

\refn\FT
{T. F\"{u}l\"{o}p and I. Tsutsui,
\plA{264}{2000}{366}}

\refn\TFCtwo
{I. Tsutsui, T. F\"{u}l\"{o}p and T. Cheon,
{\it Journ. Math. Phys.} {\bf 42} (2001) 5687}

}

%%% frontpage

%%% Output of frontpage %%%%%%%%%%%%%%%%%%%%

\pageheight{23cm}
\pagewidth{15.7cm}
\hcorrection{0mm}
\magnification= \magstep1
\def\bsk{%
\baselineskip= 16.8pt plus 1pt minus 1pt}
\parskip=5pt plus 1pt minus 1pt
\tolerance 6000

%\vsize 19.2cm   \voffset -1.2cm   %% For temporary purposes !!!

%%%%%%%%%%%%%%%%%%%%%%%%%%%%%%%%%%%%%%%%%%%%

%%% %%% %%%
% FrontPage
%%% %%% %%%

\hfill 
%\phantom
{KEK Preprint 2002-107}
\vskip -4pt 
\hfill \phantom{quant-ph/0207xxx}

\vskip 42pt

%%% Setting of the baselineskip for frontpage
{\baselineskip=18pt
%%%

\centerline{\bigbold
Supersymmetric Quantum Mechanics}
\centerline{\bigbold
with a Point Singularity}

\vskip 30pt

\centerline{\smc
Takashi Uchino\footnote{${}^*$}
{\eightpoint email:\quad uchino@post.kek.jp}
\quad
{\rm and}
\quad
Izumi Tsutsui\footnote{${}^\dagger$}
{\eightpoint email:\quad izumi.tsutsui@kek.jp}
}

\vskip 7pt

{
\baselineskip=13pt
\centerline{\it
Institute of Particle and Nuclear Studies}
\centerline{\it
High Energy Accelerator Research Organization (KEK)}
\centerline{\it Tsukuba 305-0801}
\centerline{\it Japan}
}

\vskip 100pt

\abstract{%
{\bf Abstract.}\quad
We study the possibility of supersymmetry (SUSY) 
in quantum mechanics in one
dimension under the presence of a point singularity.   
The system considered is the
free particle on a line
$\R$ or on the interval $[-l, l]$ where the point singularity lies at
$x = 0$.  In one dimension, the singularity is known to admit a $U(2)$
family of different connection conditions which include as a special case the
familiar one that arises under the Dirac delta
$\delta(x)$-potential.  Similarly, each of the walls at $x = \pm l$ admits
a $U(1)$ family of boundary conditions including the Dirichlet and the
Neumann boundary conditions.  Under these general connection/boundary
conditions, the system is shown to possess an $N = 1$ or $N = 2$ SUSY
for various choices of the singularity and the walls, and the SUSY 
is found to be \lq good\rq{} or
\lq broken\rq{} depending on the choices made.  
We use the supercharge
which allows for a constant shift in the energy, and argue that if the system
is supersymmetric then the supercharge is 
self-adjoint on states that respect the
connection/boundary conditions specified by the singularity.}

\vskip 10pt
%{\baselineskip=10pt
%{\ninepoint
%\indent{PACS codes: 3.65.-w, 2.20.-a, 73.20.Dx\hfill\break}
%\indent{Keywords: Singular interaction, Inequivalent Quantizations,
% Caustics}
%}
%}
%
%\vskip 10pt
%
%\noindent
%Number of manuscript pages: xx, \quad figures: x, \quad tables: x
%
%\centerline{
%(Running title: {\sl M{\" o}bius Structure of the Spectral Space
%under Point Interaction})
%}
%
%%% End setting of the baselineskip for frontpage
}
%%%
%\bigskip
\ve

%%% Output layout of the text %%%%%%%%%%%%%%

\pageheight{23cm}
\pagewidth{15.7cm}
\hcorrection{-1mm}
\magnification= \magstep1
\def\bsk{%
\baselineskip= 15.2pt plus 1pt minus 1pt}
\parskip=5pt plus 1pt minus 1pt
\tolerance 8000
\bsk

%%%%%%%%%%%%%%%%%%%%%%%%%%%%%%%%%%%%%%%%%%%%

\secno=1 \meqno=1

%%%%%%%%%%%%%%%%%%%%%%%%%%%%%%%%%%%%%%%%%%%%%%%%%%%%%%%%%%%%%%%%%%%%%
\bigskip
\noindent{\bf 1. Introduction}
\medskip
%%%%%%%%%%%%%%%%%%%%%%%%%%%%%%%%%%%%%%%%%%%%%%%%%%%%%%%%%%%%%%%%%%%%%

It has been known for some time that, in one dimension, quantum mechanics admits
various different singular point interactions 
parametrized by the group $U(2)$
[\RS, \AGHH].  These include the familiar singularity of
the Dirac
$\delta(x)$-potential of arbitrary strength which gives rise to discontinuity
in the derivative of the wave function, but the generic connection
condition in the $U(2)$ family develops discontinuity in both the wave
function and its derivative.   In mathematical terms, this is equivalent
to the fact that the free Hamiltonian operator, defined on the line
$\R$ with the singular point removed, admits a $U(2)$ family of
self-adjoint extensions.  
If one considers an interval $[-l, l]$ with point singularity,
then in addition to the $U(2)$ family of the singularity, the system is characterized
further by the property of the endpoints $x = \pm l$ each of which has a
$U(1)$ family of possible boundary conditions (see, {\it e.g.}, [\RS, \FCT]).  
These varieties in the connection/boundary conditions have been shown to
accommodate interesting physical phenomena, such as duality, anholonomy (Berry
phase) and scale anomaly, which are normally found in more complicated systems
or in quantum field theory [\CFT].

The varieties are also expected to furnish a room
to realize novel quantum systems with supersymmetry (SUSY).  In fact, in our
previous work [\CFT, \TFC], we found that there occurs a double degeneracy in the
energy level for some specific choices of the conditions, and an attempt was
made to reformulate the system into a SUSY quantum mechanics.  There, we
encountered the problem of 
how to ensure the self-adjointness of the supercharge
under the given connection/boundary conditions.  Another problem that needs
to be addressed is how to preserve the conditions under the transformations
generated by the supercharge.  These properties are crucial 
for the very benefit of SUSY and should be maintained, since otherwise the
generic degeneracy in the level and/or the positive semi-definiteness of the energy
will not be guaranteed.

In this paper, we provide a full analysis on the possibility of SUSY
quantum mechanics for these systems, {\it i.e.}, a free particle on the line $\R$
or on the interval $[-l, l]$ with a point singularity at $x = 0$.  We find that,
for a large variety of the connection/boundary conditions, these systems indeed
possess an $N = 1$ or
$N = 2$ SUSY (the latter case being the Witten model [\Witt]).  The
supercharge we use is a slightly extended version of the conventional one
and allows for a constant shift in the energy.  With this
supercharge, the two
properties mentioned above are shown to be maintained fully, 
if one takes the energy shift into account.   
The examples presented
include cases where the SUSY is
\lq good\rq{} or \lq broken\rq{} [\Junker], showing that these systems, though being
simple, embody the essential features of SUSY quantum mechanics observed in
other models so far.

\secno=2 \meqno=1

%%%%%%%%%%%%%%%%%%%%%%%%%%%%%%%%%%%%%%%%%%%%%%%%%%%%%%%%%%%%%%%%%%%%%
\bigskip
\noindent{\bf 2. Supersymmetry on a line with point singularity}
\medskip
%%%%%%%%%%%%%%%%%%%%%%%%%%%%%%%%%%%%%%%%%%%%%%%%%%%%%%%%%%%%%%%%%%%%%

Let us first explore the possibility of SUSY on 
a line
$\R$ in the presence of a point singularity at $x = 0$.
The system is defined by
the free Hamiltonian,
$
H = -{\hbar^2\over {2m}}\frac{d^2}{dx^2},
$
on $\R$ with the point $x = 0$ removed, and the singularity can be
characterized by a set of connection conditions at $x = 0$ for the wave
function
$\psi(x)$ belonging to
the Hilbert space ${\cal H} = L^2(\R\backslash\{0\})$.  
The system may equally be formulated by cutting the space in half (see Fig.1) and
identifying 
${\cal H}$ with $L^2(\R^+) \otimes \C^2$, where 
instead of $\psi(x)$ one considers
the vector-valued wave function, 
$$
\Psi(x) =  \left( {\matrix{{\psi_+(x)}\cr
                  {\psi_-(x)}\cr}
         }
  \right),
\qquad x \in \R^+,
\eqn\no
$$ 
defined
from
$\psi(x)$ by $\psi_+(x) = \psi(x)$ for $x > 0$ and 
$\psi_-(-x) = \psi(x)$ for $x < 0$.  
This way we introduce a
$\C^2$-graded structure into the system allowing for accommodating SUSY,
where now the Hamiltonian takes the form
$$
H = -{\hbar^2\over {2m}}\frac{d^2}{dx^2}\otimes I,
\eqn\ha
$$
with $I$ being the identity matrix acting on the $\C^2$ vector.

Before proceeding further, let us recall that the set of connection conditions
which ensures the
self-adjointness of the Hamiltonian $H$ in (\ha) is
provided by [\AG, \FT, \CFT]
$$
(U-I)\Psi(+0)+iL_0(U+I)\Psi^\prime(+0)=0.
\eqn\nyoro
$$
Here $U$, called the \lq characteristic matrix\rq{}, is an arbitrary
$U(2)$ matrix characterizing uniquely  the self-adjoint domain of $H$, which we
denote by
${\cal D}_U(H)$, and we used
$\Psi'(x) = \frac{d}{dx}\Psi(x)$.\note{%
We note that $\Psi'$ used in [\CFT] has an extra minus sign in the second
component, but this sign factor
is unnecessary here due to the mapping of the negative coordinate to the positive
one. 
}
The conditions (\nyoro) may also be written as
$$
U\Psi^{(+)}(+0) = \Psi^{(-)}(+0), \qquad \hbox{with} \quad 
\Psi^{(\pm)} = \Psi\pm iL_0\Psi^\prime.
\eqn\nyoronyoro
$$

A system is said to be \lq supersymmetric\rq{} if it has 
self-adjoint operators $Q_i$, $i = 1, \,2,\ldots$, called supercharges, such
that 
$\{Q_i, Q_j\} = H \,\delta_{ij}$ 
(see, {\it e.g.}, [\Junker]).
{}For the free Hamiltonian, the standard form of the
supercharges is
$Q_i = -i\lambda\frac{d}{dx} \otimes \sigma_i$ where $\lambda =
\hbar/\sqrt{2m}$ and $\sigma_i$ are the Pauli matrices.  Formally, these
supercharges satisfy the relation with the Hamiltonian $H$ in (\ha).  However,
this is not quite sufficient to prove that the system has a SUSY, since
operators are defined not just by the differential operations but also by the domains
on which they operate.  In fact, the supercharges $Q_i$ may not preserve the
self-adjoint domain ${\cal D}_U(H)$ of the Hamiltonian for a given $U$.  This 
can be seen, for example, by considering the domain
${\cal D}_U(H)$ for $U = \sigma_3$, for which the connection conditions read
$\psi_+^\prime(+0) = \psi_-(+0) = 0$.  The state 
$\Psi(x) = (0, xe^{-x})^T$ belongs to 
${\cal D}_U(H)$ but the transformed state 
$Q_1\Psi(x) = (i\lambda(x-1)e^{-x}, 0)^T$
do not fulfill the connection conditions and
hence $Q_1\Psi(x) \not\in {\cal D}_U(H)$.   
This problem is generic for any domain
${\cal D}_U(H)$, because the supercharges $Q_i$ involve a derivative and,
accordingly, they generate a state given basically by the derivative of the
original state.  Obviously, there is no reason to expect the generated state to
remain in the same domain as the original.  Under these
circumstances, all we can hope for is perhaps to demand that  the supercharges map
any eigenstate of the Hamiltonian $H$ to some (but not necessarily the same)
eigenstate of $H$ specified by the same
$U$.   This is reasonable because the benefit of SUSY in quantum mechanics
is that it may lead to the degeneracy of energy levels by generating an
eigenstate from a known eigenstate with the same energy by the operation of $Q_i$.   If
this is the property we want for SUSY in the system, then we do not need to require
$Q_i$ to preserve the domain ${\cal D}_U(H)$, as long as the above demand for
eigenstates is fulfilled.

\topinsert
\epsfxsize 7.0cm
\ifx\omitpictures N   \centerline{\epsfbox {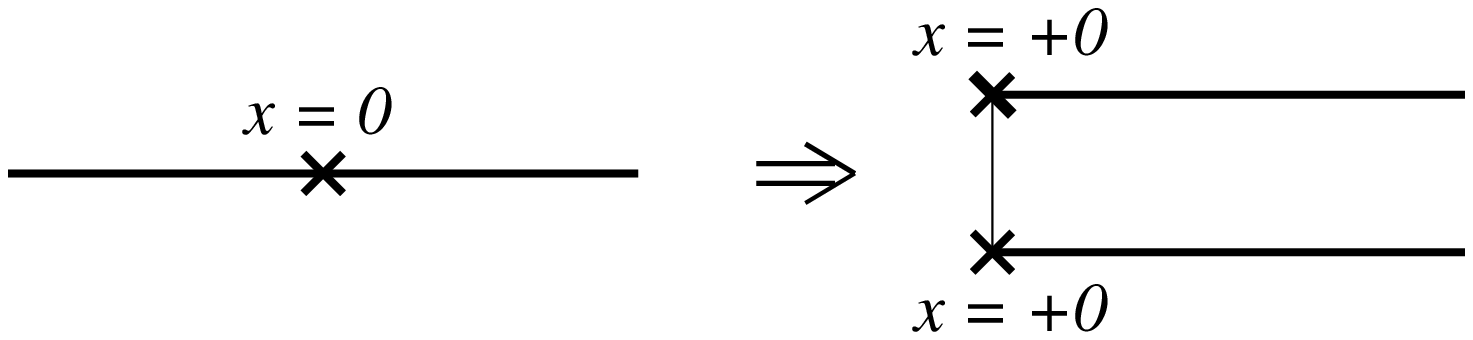}}  \fi
\abstract{{\bf Figure 1.} The system on a line $\R$ with a 
point singularity at $x = 0$
may be identified with the system of two half lines $\R^+$ with the
probability flow between the two ends $x = 0$ allowed. }
\bigskip
\endinsert

To seek for supercharges fulfilling this demand, let us consider a slightly more
general form than the standard one,
$$
Q =
-i\lambda\frac{d}{dx}\otimes\sigma_{\vec{a}}+ {\bf 1} \otimes\sigma_{\vec{b}} ,
\eqn\spp
$$
where ${\bf 1}$ is the identity operator in $L^2(\R^+)$ and
$$
\sigma_{\vec{a}} = \sum_{i = 1}^3{a_i \sigma_i},
\qquad
\sigma_{\vec{b}} = \sum_{i = 1}^3{b_i \sigma_i},
\qquad
\vert \vec{a}\vert = 1 ,
\qquad
\vec{a}\cdot \vec{b} = 0,
\qquad
a_i, \,\, b_i \in \R.
\eqn\choice
$$
The standard form is obtained if we choose $\vec{a} = (1, 0, 0)$ or $(0, 1,
0)$ and $\vec{b} = 0$. 
The properties (\choice) for the vectors $\vec{a}$ and
$\vec{b}$ lead to the relation
$2 Q^2 = H + \vert {\vec b} \vert^2$.  Now, given a domain ${\cal D}_U(H)$ of the
Hamiltonian $H$ specified by the conditions (\nyoro) with some $U$, 
our demand for $Q$ to be a supercharge  is that
$$
Q\,\Psi(x) \in {\cal D}_U(H)
\eqn\no
$$
for any $\Psi(x)$ which is an eigenstate, 
$H\,\Psi(x) = E\,\Psi(x)$,
of the Hamiltonian $H$.
If there are two (or more) such independent
operators, then one may choose an appropriate basis $Q_i$ whereby one
has
$$
\{Q_i, Q_j\} = (H+\vert {\vec b} \vert^2)\, \delta_{ij}.
\eqn\acc
$$
We note that our generalized supercharge (\spp) 
allows the constant shift $\vert {\vec b} \vert^2$ in 
(\acc) for
$ {\vec b}  \ne 0$, which is
not harmful to SUSY since the shift can always be absorbed into the Hamiltonian
by the corresponding energy shift.  It is, however, important to notice that
due to the constant in the SUSY algebra (\acc) the Hamiltonian $H$ in (\ha) may
no longer be positive semi-definite even in the presence of SUSY. 
Accordingly, the question of the system being good SUSY or broken
SUSY [\Junker] will be examined by taking this energy shift into account;
they are seen by the property of whether the supercharges annihilate the
ground state,
$Q_i\Psi^{{\rm grd}} = 0$, rather than whether a zero energy state exists or not.

To examine if a given $U$ admits such supercharges, we first note that
any $U \in U(2)$ can be decomposed as $U = V^{-1}DV$ with some matrix $V \in SU(2)$
and a diagonal matrix,
$$
D = 
	\left( \matrix{ e^{i\theta_+} & 0 \cr
	0 & e^{i\theta_-} }\right), 
\qquad \theta_\pm \in [0, 2\pi).
\eqn\diago
$$
Observe that, in view of the conditions (\nyoro) which specify the
self-adjoint domains corresponding to $U$, if
$\Psi(x)
\in {\cal D}_U(H)$ then $W \Psi(x) \in {\cal
D}_{WUW^{-1}}(H)$ for any $W \in U(2)$.  This implies that,
if there exists a pair $(U, Q)$ satisfying the above demand, so does the 
pair $(WUW^{-1}, WQW^{-1})$ where $WQW^{-1}$ is again written in the form (\spp). 
Choosing in particular
$W = V$, we find that the pair $(D, VQV^{-1})$ also satisfies the demand.  For this
reason, with no loss of generality, we restrict ourselves below 
to the case where $U$ is
diagonal.  In other words, once a solution $(D, Q)$ is found for some $D$ and
$Q$, then
$(U, V^{-1}QV)$ gives the desired solution.

To find the solutions $(D, Q)$, we observe that the charge $Q$ in (\spp) induces the
transformation on an eigenstate (with energy $E$) and its derivative as
$$
\eqalign{
\Psi(x) &\mapsto -i\lambda\sigma_{\vec{a}}\,\Psi^\prime(x)+\sigma_{\vec{b}}\,\Psi(x),
\cr
\Psi^\prime(x) &\mapsto
-i\lambda\sigma_{\vec{a}}\,\Psi^{\prime\prime}(x)+\sigma_{\vec{b}}\,
	\Psi^\prime(x) =
i\lambda^{-1}E\sigma_{\vec{a}}\,\Psi(x)+\sigma_{\vec{b}}\,\Psi^\prime(x) . 
}
\eqn\trr
$$
Our demand that the transformed state satisfy the same connection conditions
(\nyoro), or (\nyoronyoro), then implies
$$
\eqalign{
\Bigl[
\{ -\lambda(D-I) \sigma_{\vec{a}} + 2 L_0 D \sigma_{\vec{b}} \} 
&+ \{\lambda(D-I)
	\sigma_{\vec{a}} -2  L_0 \sigma_{\vec{b}} \} D \cr 
    &- L_0^2 \lambda^{-1}E (D+I) \sigma_{\vec{a}} (D + I)
\Bigr] \Psi^{(+)}(+0) = 0.
}
\eqn\senyoro
$$
An important point to be noted is that the original conditions (\nyoronyoro) provide
relations among the components between the two vectors $\Psi^{(+)}(+0)$ and
$\Psi^{(-)}(+0)$, but not among those within each of the vectors. 
This means that the equality (\senyoro) holds without the vector
$\Psi^{(+)}(+0)$.  We then see that, 
since the energy $E$ varies with the eigenstate considered, for the
conditions (\senyoro) to be identical to the original ones (\nyoro) which are
independent of $E$, the last term in the square bracket in (\senyoro) must vanish
separately from the rest.  
The conditions for SUSY therefore become
$$
(D+I) \sigma_{\vec{a}} (D+I) = 0, 
\eqn\musa
$$
and
$$
\lambda(D-I) \sigma_{\vec{a}} (D-I) 
+ 2 L_0 [D, \sigma_{\vec{b}}] = 0.
\eqn\musi
$$

{}From (\musa) one obtains
$\det(D+I) = 0$, and from (\musi) one finds $D \neq -I$.  
This shows that one of the eigenvalues of $D$ must be $-1$ while the other cannot be
$-1$.  Since the two eigenvalues in
(\diago) can be interchanged by the conjugation 
$D \rightarrow WDW^{-1}$
with
$W = i\sigma_1$, the diagonal matrix
$D$ can always be taken to be
$$
D = 
	\left( 
\matrix{ e^{i\theta} & 0 \cr
	0 & -1 }
\right),  \qquad 
\theta  \ne \pi.
\eqn\dia
$$
{}For this $D$ the condition (\musa) is fulfilled if
$$
\sigma_{\vec{a}} = \cos\alpha\, \sigma_1 + \sin\alpha\, \sigma_2 
= e^{-i{\alpha\over 2}\sigma_3}\sigma_1 e^{i{\alpha\over 2}\sigma_3},
\eqn\ndia
$$
where $\alpha \in [0, 2\pi)$ is an arbitrary angle parameter.  
If we consider in (\spp)
the simple supercharge
$Q$ with $ \vec b = 0$, then from (\musi) 
we have $\theta = 0$, {\it i.e.}, $D =
\sigma_3$ and the supercharge
$Q$ specified by the $\sigma_{\vec{a}}$ in (\ndia). 
{}For $Q$
with
$\vec
b\ne 0$, we combine (\musa) and (\musi) to find
$$
[\sigma_3, \sigma_{\vec{b}}] = {{2i\lambda}\over{L(\theta)}}\sigma_{\vec{a}},
\eqn\newcondtn
$$
where we have defined 
$$
L(\theta) = L_0 \cot{{\theta}\over 2},
\eqn\no
$$
which provides a physical length scale to the system [\TFCtwo].
The condition (\newcondtn)
can then be solved by
$$
\sigma_{\vec{b}} = {\lambda\over{L(\theta)}}
\left\{
\sin\alpha\, \sigma_1 
- \cos\alpha\, \sigma_2
\right\} + c\, \sigma_3
= -{\lambda\over{L(\theta)}} e^{-i{\alpha\over 2}\sigma_3}\sigma_2
e^{i{\alpha\over 2}\sigma_3}  + c\, \sigma_3,
\eqn\ndib
$$
with $c \in \R$ being arbitrary.  Collecting all, we find that the supercharge 
takes the form $Q = q(\alpha, c; \theta)$ with
$$
q(\alpha, c; \theta) =
-i\lambda\frac{d}{dx}\otimes
e^{-i{\alpha\over 2}\sigma_3}\sigma_1 e^{i{\alpha\over 2}\sigma_3}  +{\bf 1}
\otimes
\left[-{\lambda\over{L(\theta)}}
e^{-i{\alpha\over 2}\sigma_3}\sigma_2
e^{i{\alpha\over 2}\sigma_3}  
+ c\,
\sigma_3\right] .
\eqn\spform
$$
{}For the general $U = V^{-1}DV$, the supercharge is given by 
$$
Q = V^{-1}q(\alpha, c; \theta)V,
\eqn\gespform
$$ 
as noted above.

We therefore have learned that, if the characteristic matrix $U$ has
eigenvalues $-1$ and $e^{i\theta} \ne -1$, the system admits two independent
supercharges, {\it i.e.}, an
$N = 2$ SUSY.  For other
$U$, no SUSY is allowed under the $Q$ in (\spp).   
Since the conjugation by $V$ in (\gespform) merely rotates the
vectors in the basis $\sigma_i$, for any
$U$ that enjoys the $N = 2$ SUSY one may use the concise basis set of
supercharges,
$$
	Q_1 = -i \lambda\frac{d}{dx} \otimes \sigma_1 
- {\lambda\over L(\theta)} \otimes\sigma_2,
\qquad
	Q_2 = -i \lambda\frac{d}{dx} \otimes \sigma_2 
+ {\lambda\over L(\theta)} \otimes\sigma_1, 
\eqn\mogura
$$
by setting $\alpha =
0$ or $\alpha = \pi/2$ with $c = 0$. 
These supercharges
satisfy (\acc) with 
$\vert {\vec b} \vert^2 = [\lambda/L(\theta)]^2$.
We remark that, if we restrict ourselves to the 
simple
$Q$ with
$\vec b = 0$, then from (\musa) and (\musi) we find that an
$N = 2$ SUSY arises only if the eigenvalues of $U$ are $\pm 1$.

\secno=3 \meqno=1

%%%%%%%%%%%%%%%%%%%%%%%%%%%%%%%%%%%%%%%%%%%%%%%%%%%%%%%%%%%%%%%%%%%%%
\bigskip
\noindent{\bf 3. Supersymmetry on an interval with point singularity}
\medskip
%%%%%%%%%%%%%%%%%%%%%%%%%%%%%%%%%%%%%%%%%%%%%%%%%%%%%%%%%%%%%%%%%%%%%

Next we investigate the possibility of SUSY in the system of an
interval with a point singularity ({\it i.e.}, a quantum well with a point
interaction).  Our assumption for the supercharges remains to be of the form (\spp),
but now we need to take into account the boundary effect at both ends of the
interval.   Let the interval be $[-l, l]$ with the point singularity placed
at $x = 0$.  As before, we remove the point $x = 0$ from the interval and identify
the Hilbert space ${\cal H} = L^2([-l, l]\backslash\{0\})$ with $L^2((0, l])
\otimes \C$. The Hamiltonian (\ha) then possesses self-adjoint domains 
${\cal D}_{\tilde U}(H)$, where now the characteristic matrix $\tilde U$
belongs to
$U(2)
\times U(1)
\times U(1)$ because the point singularity at $x = 0$ furnishes a
$U(2)$ arbitrariness while each $U(1)$ corresponds to the arbitrariness 
provided by the two
ends at $x = \pm l$.  We write ${\tilde U} = U \times D_l$ where
$U \in U(2)$ is the characteristic matrix associated with $x = 0$ and  
$D_l\in U(1)\times U(1)$ is the one associated with 
the two ends
$x =\pm l$.  If we regard the two ends as a special case of point singularity 
where no
probability flow is allowed (see Fig.2), then in terms of the boundary vectors
$\Psi(l)$ and
$\Psi^\prime(l)$, the matrix $D_l$ may be taken to be a diagonal $U(1)\times
U(1)$ matrix embedded in $U(2)$.  With these, the self-adjoint domain ${\cal
D}_{\tilde U}(H)$ can be specified by the connection conditions at
$x = 0$,
$$
(U - I)\Psi(+0) + iL_0(U +I)\Psi^\prime(+0) = 0, 
\eqn\ltem
$$
together with the boundary conditions
%\note{%
%If we followed the formulation presented in [\FT], the second component of the
%boundary vectors
%$\Psi(l)$ and
%$\Psi^\prime(l)$ at the ends would have a minus sign.  The minus sign, however,  
%can always be absorbed into the matrix $D_l$ to validate (\lten).}
at $x = \pm l$,
$$
(D_l - I)\Psi(l) + iL_0(D_l +I)\Psi^\prime(l) = 0.
\eqn\lten
$$

\topinsert
\epsfxsize 7.0cm
\ifx\omitpictures N  \centerline{\epsfbox {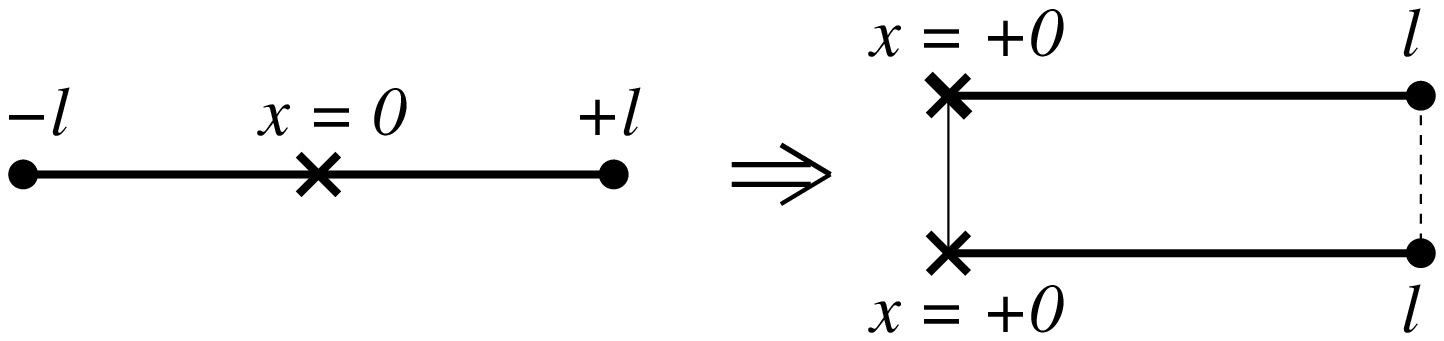}}  \fi
\abstract{{\bf Figure 2.} The system on the interval $[l,-l]$ with
a point singularity at $x = 0$ may be identified with the system of two half
intervals $(0, l]$ with the probability flow between the two ends $x = 0$ allowed.  
The flow is not allowed at the other two ends $x = l$.}
\bigskip
\endinsert

Our task is again to find a pair $({\tilde U} = U \times D_l, Q)$ 
such that the supercharge
$Q$ preserve the connection/boundary conditions (\ltem) and (\lten).
Recall that a supercharge that preserves the
conditions (\ltem) and (\lten) specified by
$U$ and $D_l$, respectively, exists only if $U$ and $D_l$ are of the form (\dia)
for $\theta \ne \pi$, up to the conjugation by some $V \in SU(2)$
for $U$.  We therefore have
$$
U = V^{-1} D V, \qquad
D =
	\left( \matrix{ e^{i\theta} & 0 \cr 
	0 & -1 } \right) ,
 \qquad
D_l =
	\left( \matrix{ e^{i\theta_l} & 0 \cr 
	0 & -1 } \right) ,
\eqn\saru
$$
for $\theta, \, \theta_l \ne \pi$, and 
our question is whether there is a supercharge $Q$ compatible to both of the
conditions (\ltem) and (\lten).  In terms of $q(\alpha, c;
\theta)$ in (\spform), the supercharge corresponding to $x = 0$ is 
$V^{-1} q(\alpha, c; \theta) V$ whereas 
the one corresponding to $x = \pm l$ is 
$q(\alpha_l, c_l; \theta_l)$, and hence our requirement of compatibility reads
$$
Q = V^{-1} q(\alpha, c; \theta) V = q(\alpha_l, c_l; \theta_l).
\eqn\supcond
$$ 
Comparing the terms involving the
derivative in (\supcond), one finds
$$
\tr\,\sigma_i \left[
V^{-1}
e^{-i{\alpha\over 2}\sigma_3}\sigma_1 e^{i{\alpha\over 2}\sigma_3}
V
\right] =\tr\, \sigma_i \left[
e^{-i{\alpha_l\over 2}\sigma_3}\sigma_1 e^{i{\alpha_l\over 2}\sigma_3}
\right],
\eqn\no
$$
for $i = 1$, 2, 3, which can be made more explicit by using the parametrization  
$$
V = e^{i{\mu\over 2}\sigma_2}e^{i{\nu\over 2}\sigma_3},
\qquad
\mu \in [0, \pi], \quad \nu \in [0, 2\pi),
\eqn\zerpo
$$
as
$$
\eqalign{
\cos\alpha \cos\mu \cos\nu 
- \sin\alpha\sin\nu &= \cos \alpha_l, \cr
\cos\alpha \cos\mu \sin\nu 
+ \sin\alpha\cos\nu &= \sin \alpha_l, \cr
\cos\alpha \sin\mu &= 0.
}
\eqn\no
$$
These are satisfied if
$$
\mu = 0, 
\qquad 
\alpha_l = \nu + \alpha,  
\qquad
\hbox{or}
\qquad
\mu = \pi, 
\qquad 
\alpha_l = \nu - \alpha \pm \pi,  
\eqn\scdc
$$
or otherwise if
$$
\alpha = {\pi \over 2}, 
\qquad 
\alpha_l = \nu + {\pi \over 2},  
\qquad
\hbox{or}
\qquad
\alpha = {{3\pi} \over 2}, 
\qquad 
\alpha_l = \nu - {\pi \over 2}. 
\eqn\fstc
$$

The remaining conditions which arise from 
the non-derivative term in (\supcond) are
$$
\eqalign{
&\tr\, \sigma_i  
\left[ 
-{\lambda\over{L(\theta)}}
V^{-1}e^{-i{\alpha\over 2}\sigma_3}\sigma_2
e^{i{\alpha\over 2}\sigma_3} V
+ c\,
V^{-1}\sigma_3 V
\right]  \cr
&\qquad\qquad =
\tr\, \sigma_i 
\left[ 
-{\lambda\over{L(\theta_l)}}
e^{-i{\alpha_l\over 2}\sigma_3}\sigma_2
e^{i{\alpha_l\over 2}\sigma_3}  
+ c_l\,
\sigma_3
\right].
}
\eqn\nondercond
$$
Among these conditions the one corresponding to the third component $i = 3$ can
always be fulfilled by adjusting the free constants $c_l$ and $c$. 
Also, only
one of the remaining two components, say $i = 2$, is important because, once
its corresponding condition is met, the construction of the
supercharge ensures that the other 
component $i = 1$ fulfills its condition, too.   The
$i = 2$ component of (\nondercond) is
$$
-{\lambda\over{L(\theta)}}
\left(\cos\alpha \cos\nu
- \sin\alpha \cos\mu \sin\nu 
\right)  + c \sin\mu \sin\nu =
-{\lambda\over{L(\theta_l)}} \cos \alpha_l.
\eqn\scscdcond
$$

Now, for (\scdc) we observe that (\scscdcond)
simplifies into $L(\theta) = \pm L(\theta_l)$ or $\theta = \pm \theta_l$.
We thus realize that, if the point singularity at $x = 0$
and the endpoints at $x = \pm l$ are characterized by 
$$
U =
	\left( \matrix{ e^{i\theta} & 0 \cr 
	0 & -1 } \right) ,
 \qquad
D_l =
	\left( \matrix{ e^{i\theta} & 0 \cr 
	0 & -1 } \right) ,
\qquad
\theta \ne \pi,
\eqn\csone
$$
or
$$
U =
	\left(\matrix{ -1 & 0 \cr 
	0 &  e^{-i\theta}} \right) ,
 \qquad
D_l =
	\left( \matrix{ e^{i\theta} & 0 \cr 
	0 & -1 } \right) ,
\qquad
\theta \ne \pi,
\eqn\cstwo
$$
the system has the supercharge
preserving the connection/boundary conditions simultaneously.  The supercharge
is 
$$
Q = q(\alpha, c; \theta) ,
\eqn\fstspc
$$
where $q(\alpha, c; \theta)$ is given in (\spform).  Since the angle parameter
$\alpha$ in the derivative term in (\fstspc)
remains arbitrary, we see that the system admits an
$N = 2$ SUSY.

On the other hand, for (\fstc) we observe that 
(\scscdcond) reduces to $\nu = 0$, $\pi$, or
$$
\pm {\lambda\over{L(\theta)}}
\cos\mu 
+ c \sin\mu  =
\pm {\lambda\over{L(\theta_l)}}.
\eqn\no
$$
This relation may be used to determine the constant $c$ in favor of 
$\mu$, $\theta$ and $\theta_l$.  
The relevant supercharge then reads
$$
Q = q(\nu \pm \pi/2, c; \theta) ,
\eqn\scdspc
$$
where $\nu$ is one of the 
parameters in (\zerpo).  In contrast to (\fstspc), 
the angle parameter in (\scdspc) is determined by $\nu$ in $U$, and hence
the system admits only an
$N = 1$ SUSY.

To see in more detail the content of the SUSY systems we have
found,  we consider, for instance, the case (\csone) whose 
connection/boundary conditions are 
$$
\eqalign{
\psi_+(+0) +L(\theta)\, \psi_+^\prime(+0) &= 0, 
\qquad \psi_-(+0) = 0, \cr
\psi_+(l) +L(\theta)\, \psi_+^\prime(l) &= 0, 
\qquad \psi_-(l) = 0.
}
\eqn\cbone
$$
The eigenstates of the Hamiltonian $H$ with positive energy 
$E_n = \hbar^2k_n^2/(2m) > 0$ satisfying (\cbone) are doubly degenerate,
$$
\Psi^{(n)}_+(x) =
		N_+^{(n)}\left( \matrix{ -L(\theta)\,k_n \cos k_n x + \sin k_n x \cr
		0 } \right), 
\qquad
\Psi^{(n)}_-(x) =
		N_-^{(n)}\left( \matrix{ 0 \cr
		\sin k_n x } \right),
\eqn\no
$$
where $k_n = n\pi/l$ with $n = 1,2,3,\ldots$, and 
$N_+^{(n)}$ and $N_-^{(n)}$ 
%$N_+^{(n)} = [l\{(L(\theta)k_n)^2 + 1\}/2]^{-1/2}$ and
%$N_-^{(n)} = (l/2)^{-1/2}$ 
are normalization constants. 
Under the SUSY transformations  generated by (\mogura), these two states
interchange each other, 
$\Psi^{(n)}_\pm \mapsto Q_i\Psi^{(n)}_\pm \propto \Psi^{(n)}_\mp$
for $i = 1, 2$.
The ground state is given by
$$
\Psi^{{\rm grd}}_+(x) = 
N_+^{{\rm grd}}	\left( \matrix{ e^{-x/L(\theta)} \cr
	0 } \right),
\eqn\ora
$$
with energy 
$E^{\rm grd} = - \hbar^2/\{2m(L(\theta))^2\} 
= -[\lambda/ L(\theta)]^2 < 0$, 
% and $N^{{\rm grd}} = [L(\theta)(1 - e^{-2l/L(\theta)})/2]^{-1/2}$, 
which
exists for any
$\theta \ne \pi$ 
except when $\theta = 0$ which yields the Neumann condition,
$\psi_+'(+0) = \psi_+'(l) = 0$.  The ground state is unique and annihilated by the
supercharge 
$Q_i\Psi_+^{{\rm grd}} = 0$, which shows that the system possesses a good SUSY.

{}For the case (\cstwo), on the other hand, the 
connection/boundary conditions read 
$$
\eqalign{
\psi_+(+0) &= 0, 
\qquad \psi_-(+0) -L(\theta)\, \psi_-^\prime(+0) = 0,
\cr 
\psi_+(l) +L(\theta)\, \psi_+^\prime(l) &= 0, 
\qquad \psi_-(l) = 0.
}
\eqn\cbtwo
$$
The eigenstates with positive energy are then
$$
\Psi^{(n)}_+(x) =
N_+^{(n)}	\left( \matrix{ \sin k_n x \cr
    0 } \right) ,
\qquad
\Psi^{(n)}_-(x) =
	N_-^{(n)}\left( \matrix{ 0 \cr
    \sin k_n(x-l) } \right) ,
\eqn\no
$$
where the discrete $k_n > 0$ are determined as solutions of 
$L(\theta) k_n + \tan(k_n l) = 0$.  Similarly, 
there arise the ground states,
$$
\Psi^{{\rm grd}}_+(x) =
N_+^{{\rm grd}}	\left( \matrix{ \sinh \kappa x \cr
    0 } \right) ,
\qquad
\Psi^{{\rm grd}}_-(x) =
	N_-^{{\rm grd}}\left( \matrix{ 0 \cr
    \sinh \kappa (x-l) } \right) ,
\eqn\no
$$
with $E^{{\rm grd}} = -\hbar^2\kappa^2/(2m) < 0$, 
where $\kappa > 0$ satisfies 
$L(\theta) \kappa + \tanh \kappa l = 0$ which has a solution for
$-l < L(\theta) < 0$.  If $L(\theta) = -l$, we have in addition the 
zero energy states,
$$
\Psi^{{\rm zero}}_+(x) =
N_+^{{\rm zero}}	\left( \matrix{ x \cr
    0 } \right) ,
\qquad
\Psi^{{\rm zero}}_-(x) =
	N_-^{{\rm zero}}\left( \matrix{ 0 \cr
    x-l } \right) .
\eqn\no
$$
Irrespective of the energy (positive, negative or zero), all states are doubly
degenerate and the degenerate pair of states are related by the SUSY
transformations generated by $Q_i$.  The degeneracy of the ground states implies
that, in contrast to the previous case, the SUSY is broken here.  Note that the
energy of the ground states does not attain the lower bound,
$E^{{\rm grd}} > - [\lambda/ L(\theta)]^2$.

In particular, if we choose $\theta = 0$, then the conditions (\cbone) become
the Neumann type $\psi_+^\prime(+0) = 0 = \psi_+^\prime(l)$ and the 
Dirichlet type $\psi_-(+0) = 0 = \psi_-(l)$, whereas the
conditions (\cbtwo) become their combinations, $\psi_+(+0) =
0 =\psi_+^\prime(l)$ and $\psi_-^\prime(+0) = 0 = \psi_-(l)$.
These are the models known earlier [\Junker] whose supercharges are 
given without the constant term in (\spp).  
Taking into account the $N = 1$ systems realized 
on the interval under (\scdspc), we
see that, under the supercharge (\spp) with a constant term,  the general point
singularity leads to 
a much richer variety of SUSY systems than known before.  

If we restrict ourselves to the 
simple $Q$ with $\vec b = 0$, then from the result in section 2, we know
that both of
$U$ and $D_l$ must have the eigenvalues $+1$ and $-1$, that is, 
$U = V^{-1}\sigma_3 V$ and $D_l = \sigma_3$ (up to the exchange of the
eigenvalues). This implies the connection/boundary conditions,
$$
\eqalign{
e^{i\nu} \psi_+(+0) - \cot {\mu\over 2}\, \psi_-(+0) &= 0, 
\qquad e^{i\nu} \psi_+^\prime(+0) + \tan {\mu\over 2}\, 
\psi_-^\prime(+0) = 0,
\cr 
\psi^\prime_+(l) &= 0, 
\qquad 
\psi_-(l) = 0,
}
\eqn\cbtwo
$$
under which we have the eigenstates,
$$
\Psi^{(n)}(x) = N^{(n)}	 
		\left( \matrix{ - e^{-i\nu} \cos k_n (x-l) \cr
        \sin k_n (x-l) } \right) ,
\qquad
k_n = {{n\pi + \mu/2}\over{l}},
\eqn\no
$$
for $n \in \Z$.  Each eigenstate is invariant
under the SUSY transformation by $Q$, and as shown in Fig.3,
the energy levels $E^{(n)} = \hbar^2 k_n^2/(2m)$ 
are not degenerate unless $\mu = 0$ or $\pi$.

\topinsert
\epsfxsize 5.0cm
\ifx\omitpictures N  \centerline{ \epsfbox {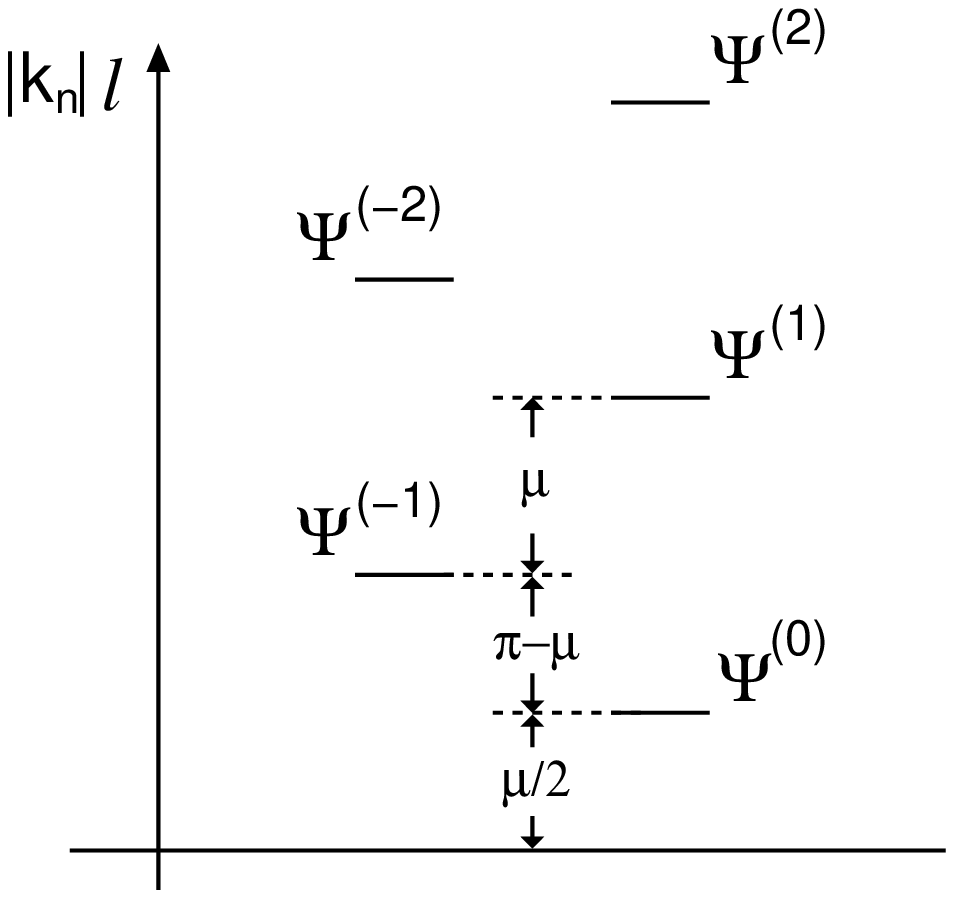} }  \fi
\abstract{{\bf Figure 3.}~Energy levels of the $N = 2$ SUSY system
possessing the simple supercharge $Q$ with $\vec b = 0$.  The levels
are not degenerate unless $\mu = 0$ or $\pi$.}
\bigskip
\endinsert

{}For the interval we have yet another extension based on the \lq half
parity\rq{} transformation,
$$
{\cal X}: \Psi(x) \mapsto  
({\cal X}\Psi)(x) =  \left( {\matrix{{\psi_+(l - x)}\cr
                  {\psi_-(x)}\cr}
         }
  \right),
\eqn\hprty
$$
which is well-defined 
in ${\cal H}$ and fulfills ${\cal X}^2 = \hbox{id}$.  For the
system characterized by $\tilde U = U \times D_l$ with $U$ and
$D_l$ given either by (\csone) or (\cstwo), this half parity
induces a change in the 
characteristic matrix $\tilde U \mapsto \tilde U^{\cal X} = U^{\cal X} \times
D_l^{\cal X}$, namely, if $\Psi \in {\cal D}_{\tilde U}(H)$ then ${\cal
X}\Psi \in {\cal D}_{\tilde U^{\cal X}}(H)$, where $U^{\cal X}$ and 
$D_l^{\cal X}$ are given by interchanging the upper left components of
$U$ and $D_l$ with the extra sign $\theta \rightarrow  -\theta$ . 
Clearly, we have  
$Q ({\cal X}\Psi) \in {\cal D}_{\tilde U^{\cal
X}}(H)$ for the supercharge 
$Q = {\cal X} q(\alpha, c; \theta) {\cal X}^{-1}$. 
The half parity transformation ${\cal X}$ leaves the good SUSY system 
(\csone) unaltered, but it turns the broken SUSY system (\cstwo) into
one with
$$
U^{\cal X} =
	\left(\matrix{ e^{-i\theta} & 0 \cr 
	0 &  e^{-i\theta}} \right) ,
 \qquad
D^{\cal X}_l =
	\left( \matrix{ - 1 & 0 \cr 
	0 & -1 } \right) ,
\qquad
\theta \ne \pi.
\eqn\csthree
$$
Under these the connection/boundary conditions read
$$
\eqalign{
\psi_+(+0) -L(\theta)\, \psi_+^\prime(+0) &= 0, 
\qquad \psi_-(+0) -L(\theta)\, \psi_-^\prime(+0) = 0,
\cr 
\psi_+(l) &= 0,
\qquad \psi_-(l) = 0.
}
\eqn\exbb
$$
Evidently, this system has a spectrum identical to the case (\cstwo) and
the $N = 2$ SUSY is broken.   The present case corresponds to
the \lq self-dual\rq{} subfamily mentioned earlier in Ref.[\TFC], which 
pointed out that the system becomes a Witten model at $\theta = \pi$ but 
fell short of obtaining the full realization of SUSY quantum mechanics 
for other $\theta$ due to the
question of the self-adjointness of the supercharge, which we now address
next.

\secno=4 \meqno=1

%%%%%%%%%%%%%%%%%%%%%%%%%%%%%%%%%%%%%%%%%%%%%%%%%%%%%%%%%%%%%%%%%%%%%
\bigskip
\noindent{\bf 4. Self-Adjointness of the supercharge}
\medskip
%%%%%%%%%%%%%%%%%%%%%%%%%%%%%%%%%%%%%%%%%%%%%%%%%%%%%%%%%%%%%%%%%%%%%

An important question which remains to be answered is
whether our supercharge
$Q$ in (\gespform) can be defined as a self-adjoint operator, and if so,
whether  its self-adjoint domain 
is compatible with the self-adjoint domain of the
Hamiltonian.  

To answer the former half of the question, we first observe that the problem of
the self-adjointness can be examined by the form (\spp) because, analogously
to the case of the Hamiltonian, a self-adjoint domain for (\gespform), if
any, can be obtained from a domain for (\spp) by the conjugation of the matrix
characterizing the domain of the supercharge.  Now, for the 
interval $[-l, l]$, for example, the supercharge
(\spp) being self-adjoint implies
$$
\int_{+0}^l\Psi^\dagger(x) (Q\Psi)(x) dx 
- \int_{+0}^l(Q\Psi)^\dagger(x) \Psi(x)
dx = 0.
\eqn\marimo
$$
Clearly, the constant term in $Q$ drops out from this
condition and hence does not affect the domain determined from
(\marimo).  
Taking
the freedom of conjugation (including the half parity
${\cal X}$, if necessary) into account, with no loss of generality we
can restrict ourselves to the simple form 
$Q = -i\lambda\frac{d}{dx}\otimes\sigma_2$.  Then, the
condition (\marimo) reduces to
$$
\Psi^\dagger(l)\sigma_2\Psi(l) - 
\Psi^\dagger(+0)\sigma_2\Psi(+0)= 0.
\eqn\yago
$$
This can be fulfilled if
$\Psi(0) = M\Psi(l)$ with some $U(1, 1)$ matrix $M$, which satisfies
$M^\dagger \sigma_2 M = \sigma_2$.  Other solutions are given by
$\psi_+(+0) + u\,\psi_-(+0) = 0$ and
$\psi_+(l) + u_l\,\psi_-(l) = 0$, where 
$u, u_l \in \R \cup \{\infty\} \simeq U(1)$. 
The entire family of the solutions is given by the sum of these, and
they actually form a $U(2)$ group.  That this is the case may be argued 
by invoking the theory of self-adjoint extensions [\RS] as follows.  

Let ${\cal D}(Q)$ be a symmetric domain of 
$Q$ defined by 
$$
{\cal D}(Q)= \left\{\Psi\, \big|\, \Psi\in AC([0,l])\otimes\C^2, \,  
Q\Psi\in L^2((0,l])\otimes\C^2,\,
\Psi(+0) = \Psi(l) = 0 \right\},
\eqn\no
$$
where $AC([0,l])$ denotes the space of absolutely continuous functions on
the interval $[0,l]$.  
The domain of the adjoint $Q^\dagger$ is then given by
$$
{\cal D}(Q^\dagger)= \left\{\Psi\, \big|\, 
\Psi\in AC([0,l])\otimes\C^2, \, 
Q\Psi\in L^2((0,l])\otimes\C^2 \right\}.
\eqn\no
$$
The eigenvalue equation,
$Q\Psi_{\pm i} = (-i\lambda\frac{d}{dx}\otimes\sigma_2)\Psi_{\pm i} = \pm
i g\, \Psi_{\pm i}$ for any $g > 0$, has the solutions,
$$
\Psi^{(1)}_{\pm i} =
	\left( \matrix{ \pm i \cr 1 } \right) e^{g x/\lambda} , 
\qquad
\Psi^{(2)}_{\pm i} = 
	\left( \matrix{ \pm i \cr -1 } \right) e^{- g x/\lambda} .
\eqn\eifun
$$
Thus the deficiency indices of the operator $Q$ are $(2, 2)$ showing that 
the supercharge $Q$ admits a $U(2)$ family 
of self-adjoint extensions for the interval.
These extensions are characterized by the aforementioned boundary conditions.

{}For the line $\R$, the above discussion applies more or less unchanged, except
that the contribution from the endpoint
$x = l$ is now absent.  The deficiency indices of the operator $Q$ become $(1,
1)$ since $\Psi^{(1)}_{\pm i}$ are non-normalizable there.  The resultant 
$U(1)$ family of self-adjoint domains is realized by the
boundary condition,
$\psi_+(+0) + u\,\psi_-(+0) = 0$.
 
We next turn to the latter half of the question, namely, the compatibility of
the two self-adjoint domains of $Q$ and $H$.  This, again, can be answered
affirmatively.  In fact, we show that any self-adjoint domain ${\cal D}_U(H)$
is a subset of the self-adjoint domain of the supercharge $Q$ associated with
the Hamiltonian $H$.  To see this,
let $f_k$, $k = 1, 2$ be the eigenvectors of the characteristic matrix $U$
specifying the Hamiltonian.  We then have 
$Uf_k =e^{i\theta_i}f_k$, where one of the eigenvalues, say $e^{i\theta_2}$, is
$-1$ while the other $e^{i\theta_1}$ is not $-1$.  We now decompose the
boundary vectors
$\Psi(+0)$ and
$\Psi^\prime(+0)$ into the eigenvectors of the
characteristic matrix $U$ of the Hamiltonian,
$$
\Psi(+0)= \sum_{k = 1}^{2}{\langle f_k, \Psi(+0)\rangle \, f_k}, \qquad
\Psi^\prime(+0) = \sum_{k = 1}^{2}{\langle f_k, \Psi^\prime(+0)\rangle
\,f_k}, 
\eqn\no
$$
where $\langle \cdot, \cdot \rangle$ is the innerproduct for $\C^2$-vectors. 
In terms of these, the connection conditions (\nyoro) become
$$
\sum_{k=1}^{2}{\bigl[ (e^{i\theta_i}-1)\langle f_k, \Psi(+0)\rangle \,
+ iL_0(e^{i\theta_i}+1) \langle f_k, \Psi^\prime(+0)\rangle \bigr] f_k } = 0.
\eqn\mizusumasi
$$
From the independence of the eigenvectors $f_k$ and $e^{i\theta_2} = -1$, we
find
$$
\langle f_2, \Psi(+0)\rangle = 0.
\eqn\yamori
$$
On the other hand, we recall that 
the existence of the supercharge requires (\musa).  For $\sigma_{\vec{n}}$
now taken by $\sigma_2$, and for general $U$, (\musa) becomes 
$(U+I) \sigma_2 (U+I) = 0$. 
Multiplying this by $f_1$ from the right and by
$f_1^\dagger U^\dagger$ from the left, we obtain
$$
f_1^\dagger\,\sigma_2\, f_1 = 0.
\eqn\tanisi
$$
From (\yamori) and (\tanisi), we see that
$$
\Psi^\dagger(+0)\sigma_2\Psi(+0) =
\sum_{j,k = 1}^{2}{\langle f_k, \Psi(+0)\rangle 
\langle \Psi(+0), f_j\rangle
\,f_j^\dagger\sigma_2 f_k} = 0.
\eqn\yadokari
$$
So far we have considered only for $x  = +0$, but 
the contribution from $x = l$ can be evaluated analogously to show that
$\Psi^\dagger(l)\sigma_2\Psi(l) = 0$.  We therefore realize that the
requirement for the self-adjointness (\yago) of $Q$ is ensured 
for any states belonging to the domain of $H$ for which an associated
supercharge
$Q$ exists, and that this is true for both the line and the
interval system.

An important consequence of this is that all eigenvalues of $Q$ are real
and, hence, the operator $Q^2$ is positive semi-definite.  From this
we find the lower bound of the spectrum,
$H = 2Q^2 - [\lambda/L(\theta)]^2\otimes {\bf 1} \ge -
[\lambda/L(\theta)]^2\otimes {\bf 1}$, which is attained by the ground state
(\ora) in the good SUSY case.

\secno=5 \meqno=1

%%%%%%%%%%%%%%%%%%%%%%%%%%%%%%%%%%%%%%%%%%%%%%%%%%%%%%%%%%%%%%%%%%%%%
\bigskip
\noindent{\bf 5. Conclusion and discussions}
\medskip
%%%%%%%%%%%%%%%%%%%%%%%%%%%%%%%%%%%%%%%%%%%%%%%%%%%%%%%%%%%%%%%%%%%%%

We have seen that a rich variety of $N = 1$ and $N = 2$ SUSY systems appear
on the line and the interval with a singular point.  The key element for this
is that we consider the entire family of quantum singularities, and that we
extend the supercharge by introducing a constant term allowing for a shift in
the energy.  The resultant $N = 2$ SUSY systems are the Witten models
and exhibit different features depending on the choice of the characteristic
matrix that specifies the singularity.  

The self-adjoint domain of the
supercharge $Q$ is seen to contain the self-adjoint domain of the Hamiltonian
$H$.  As a result, the supercharge $Q$ may be expected to ensure the double
degeneracy of energy levels by its operation on the eigenstates.  Indeed, this
has been seen in the first two (and the fourth) examples discussed 
in the interval case in section 3, but
not in the third one where no degeneracy arises in general.  The standard tool to
establish the degeneracy is the Witten parity operator $W$, which is
self-adjoint and satisfies
$W^2= {\bf 1}$, $[W, H]=0$ and $\{W, Q\}=0$.
Obviously, for these conditions we need to examine if the domains of the
operators invloved --- in particular the
domain ${\cal D}_{\tilde U}(H)$ of the Hamiltonian --- change 
under the operation of
$W$, and this is a highly nontrivial matter.  However, if we assume that 
$W$ is given by a $2\times 2$ Hermitian matrix acting on the
graded Hilbert space, then on account of the boundary conditions (\nyoro) we see
that
${\cal D}_{\tilde U}(H)$ is preserved  under $W$ if $[W, U] = [W, D_l] = 0$.   
The first two (and the fourth) examples have their ${\tilde U}$ 
that fulfills this demand with 
$W = \sigma_3$, and consequently allow the degeneracy to occur in the
eigenspaces of $\sigma_3$, {\it i.e.}, the upper and the lower components of
the vector states $\Psi$.  In contrast, in the
third example, $U$ and $D_l$ do not in general allow a common operator commuting
the both simultaneously, implying that such $W$ cannot exist.  In fact, this
seems to be the case for a generic pair of $U$ and $D_l$ unless there underlies 
some mechanism to ensure the degeneracy.

{}Finally, we mention that it is straightforward to extend our analysis to more
complicated systems with point singularities, including a circle with singular
points or a quantum circuit whose vertices may be regarded as point
singularities.  For these systems, all we need is to put an appropriate
$U(2)$ matrix to each of the point singularity and seek the supercharge that
preserves the domain of the Hamiltonian.  
The extension may also include systems with a
potential $V(x)$ which develops a singularity at its divergent point, such as the
Coulomb potential.  Singular potentials that arise in integrable models,
such as the Calogero-Moser models, may also be of interest for the possibility
of accommodating SUSY under the general singularity.

\bigskip
\noindent
{\bf Acknowledgement:}
I.T.~thanks T. Cheon and T. F\"{u}l\"{o}p for useful discussions.
This work has been supported in part by
the Grant-in-Aid for Scientific 
Research on Priority Areas (No.~13135206) by
the Japanese
Ministry of
Education, Science, Sports and Culture.

\baselineskip= 15.5pt plus 1pt minus 1pt
\parskip=5pt plus 1pt minus 1pt
\tolerance 8000
\vfill\eject\immediate\closeout\reffile%\parindent=20pt
\centerline{{\bf References}}\bigskip\frenchspacing%
\input refs.tmp\vfill\eject\nonfrenchspacing

\bye